\documentclass[preprint,prd,
amsmath,amssymb,amsthm,nofootinbib,superscriptaddress,longbibliography]{revtex4-2}

\usepackage{graphicx}   
\usepackage[
colorlinks=true,        
citecolor=blue,         
linkcolor=blue,         
urlcolor=blue ,          
]{hyperref}  
\usepackage{color}      
\usepackage{orcidlink}
\usepackage{multirow}
\usepackage{bm}
\usepackage[stretch=10]{microtype}

\newcommand{\nc}{\newcommand*} 
\newcommand{\be}{\begin{equation}}
	\newcommand{\ee}{\end{equation}}
\newcommand{\bea}{\setlength\arraycolsep{2pt} \begin{eqnarray}}
	\newcommand{\eea}{\end{eqnarray}}
\newcommand{\nn}{\nonumber}
\newcommand{\de}{{\rm d}}
\nc{\al}{\alpha}
\nc{\s}{\sigma}
\nc{\dt}{\delta}
\nc{\Dt}{\Delta}
\nc{\Ld}{\Lambda}
\nc{\p}{\partial}
\nc{\om}{\omega}
\nc{\Om}{\Omega}
\nc{\rd}{\mathrm{d}}
\nc{\Od}[1]{\mathcal{O}(#1)} 
\nc{\kp}{\kappa}

\def\e{\begin{equation}}
	\def\q{\end{equation}}
\def\m{\begin{eqnarray}}
	\def\n{\end{eqnarray}}

\nc{\Eq}[1]{Eq.~\eqref{#1}}     
\nc{\Fig}[1]{Fig.~\ref{#1}}     
\nc{\Table}[1]{Tab.~\ref{#1}}  
\nc{\Sec}[1]{Sec.~\ref{#1}}     

\nc{\Msun}{M_\odot}             
\nc{\fpbhn}{f_{\mathrm{pbh0}}}    
\nc{\mR}{\mathcal{R}} 
\nc{\seq}{\sigma_{\mathrm{eq}}}
\nc{\ogw}{\Omega_{\mathrm{GW}}}
\nc{\gpcyr}{\mathrm{Gpc}^{-3}\,\mathrm{yr}^{-1}}
\nc{\lvc}{LIGO/Virgo} 
\nc{\SNR}{\mathrm{SNR}} 
\nc{\mmin}{{m_{\mathrm{min}}}}
\nc{\mmax}{{m_{\mathrm{max}}}}
\nc{\Mmin}{{M_{\mathrm{min}}}}
\nc{\fmin}{{f_{\mathrm{min}}}}
\nc{\VT}{\mathrm{VT}}
\nc{\rhoGW}{\rho_{\mathrm{GW}}}
\nc{\vth}{\vec{\theta}}
\nc{\vd}{\vec{d}}
\nc{\vla}{\vec{\lambda}}
\nc{\bLa}{\bm{\Lambda}}
\nc{\Nobs}{N_{\mathrm{obs}}}
\nc{\av}[1]{\langle #1 \rangle} 
\nc{\km}{\mathrm{km}}
\nc{\Mpc}{\mathrm{Mpc}}
\nc{\Tobs}{T_{\mathrm{obs}}}
\nc{\Ntemp}{N_{\mathrm{temp}}}
\nc{\ie}{\textit{i.e.}}
\nc{\eg}{\textit{e.g.~}}
\nc{\app}{\approx}
\nc{\hf}{\frac{1}{2}}

\newcommand{\apjl}{Astrophys. J. Lett.}

\newcommand{\aap}{Astron. \& Astrophys.}

\newcommand{\araa}{Ann. Rev. Astron. Astrophys.} 
\newcommand{\mnras}{Mon. Not. R. Astron. Soc.}

\newcommand{\pasa}{Pub. Astro. Soc. Aust.}

\newcommand{\physrep}{Physics Reports}

\def \kmsMpc {\mathrm{km\ s^{-1}\ Mpc^{-1}}}

\begin{document}

\title{Probing the Mass--Redshift Dependence of Binary Black Holes and its Implications for $H_0$ with GWTC-5.0}

\author{Yang Liu }
\email{yangliu@pmo.ac.cn}
\affiliation{Purple Mountain Observatory, Chinese Academy of Sciences, No. 10 Yuanhua Road, Nanjing 210023, China}

\author{Puxun Wu}
\email{pxwu@hunnu.edu.cn}
\affiliation{Department of Physics, Key Laboratory of Low Dimensional Quantum Structures and Quantum Control of Ministry of Education, and Hunan Research Center of the Basic Discipline for Quantum Effects and Quantum Technologies, Hunan Normal University, Changsha, Hunan 410081, China}

\author{Jun-Jie Wei}
\email{jjwei@pmo.ac.cn}
\affiliation{Purple Mountain Observatory, Chinese Academy of Sciences, No. 10 Yuanhua Road, Nanjing 210023, China}

\author{Xue-Feng Wu}
\email{xfwu@pmo.ac.cn}
\affiliation{Purple Mountain Observatory, Chinese Academy of Sciences, No. 10 Yuanhua Road, Nanjing 210023, China}

\begin{abstract}

The mass and redshift distributions of merging binary black holes (BBHs) bear imprints of their astrophysical formation channels. Whether the BH mass distribution evolves with redshift, however, remains an open question. In this paper, we employ copula functions, which describe the dependence between variables independently of their marginal distributions, to probe the dependence structure between the primary mass and the redshift of BBH mergers. We construct five population models by coupling the marginal distributions of primary mass and redshift with the Gaussian copula, the Clayton copula, or one of its three rotations, and we constrain their hyperparameters using 235 BBH events from GWTC-5.0. For comparison, we also adopt a baseline model in which the primary mass and redshift are assumed independent. Among the copula-based models, only the model with the $180^\circ$ rotated Clayton copula, which couples higher primary masses preferentially with high redshifts, exhibits evidence for a nonzero correlation, with its copula parameter deviating from the independence limit at more than $1\sigma$ significance. 
Importantly, the constraints on the Hubble constant ($H_0$) derived from these models are mutually consistent within the $68\%$ confidence level; the baseline model yields $H_0=74.3^{+13.7}_{-19.6}~\kmsMpc$, indicating that allowing for a mass--redshift dependence does not significantly bias the $H_0$ inference with the current dataset. 
Bayesian model comparison favors the baseline model over all copula-based alternatives, with log-Bayes factors corresponding to weak to moderate evidence against the latter on the Jeffreys scale. Our results thus imply that the current GWTC-5.0 sample remains consistent with no redshift evolution of the primary mass distribution over the redshift range probed.

\end{abstract}
\maketitle

\section{Introduction}

Since the first direct detection of gravitational-waves (GWs) from the binary black hole (BBH) merger GW150914~\cite{2016PhRvL.116f1102A}, the LIGO--Virgo--KAGRA (LVK) detector network has completed the second part of its fourth observing run (O4b). The fifth Gravitational-Wave Transient Catalog (GWTC-5.0) now contains over 300 compact binary coalescence (CBC) candidates, including 161 new events from this run~\cite{2026arXiv260527225T}. 
The detected population encompasses binary neutron star (BNS) mergers~\cite{2017PhRvL.119p1101A,2020ApJ...892L...3A}, neutron star-black hole (NS-BH) mergers~\cite{2021ApJ...915L...5A}, and, most abundantly, BBH mergers.
These signals provide measurements of the component masses, spins, and
luminosity distances of the sources, thereby enabling detailed studies of their
population properties~\cite{2021ApJ...913L...7A,2023PhRvX..13a1048A,2026arXiv260527226T}. 

Population analyses of these GW events have revealed rich features in the BBH mass distribution. The primary-mass distribution deviates from a simple power law, exhibiting a robust peak near $\sim 10\,\Msun$ and a change of slope around $\sim 35\,\Msun$~\cite{2026arXiv260527226T}. 
Above this scale the distribution steepens and extends smoothly beyond $100\,\Msun$~\cite{2023PhRvX..13a1048A,2026ApJ..1005L..51A,2026arXiv260527226T}.
These features may encode information about the formation and evolution of compact binaries. The $\sim 10\,\Msun$ peak is consistent with expectations from isolated binary evolution~\cite{2015ApJ...806..263D, 2018MNRAS.480.2011G, 2019ApJ...885....1W, 2019MNRAS.490.3740N, 2023ApJ...948..105V, 2022ApJ...940..184V}, whereas
the origin of the $\sim 35\,\Msun$ feature remains less settled. 
It was initially interpreted as a pile-up produced by pulsational pair-instability supernovae~\cite{2016A&A...594A..97B, 2017ApJ...836..244W, 2017MNRAS.470.4739S, 2019ApJ...882..121S, 2018ApJ...856..173T,2019ApJ...887...53F, 2019ApJ...882...36M, 2025MNRAS.540...90W}. However,
placing this feature at such a low mass requires the $^{12}\mathrm{C}(\alpha,\gamma)^{16}\mathrm{O}$ reaction rate to exceed its laboratory-measured value, so its astrophysical interpretation remains open~\cite{2020ApJ...902L..36F,2021ApJ...912L..31W, 2023MNRAS.526.4130H, 2025PhRvD.112f3053C, 2024ApJ...976..121G, 2025CQGra..42v5008K}. Moreover, no clear signature of the pair-instability mass gap has emerged in the observed population~\cite{2026arXiv260527226T}, and a few events, such as GW190521~\cite{2020PhRvL.125j1102A,2020ApJ...900L..13A}, even lie within the predicted gap.

Joint analyses of mass with other observables help disentangle the formation channels of BBHs~\cite{2022PhR...955....1M,2020FrASS...7...38M}.
Studies combining mass and spin distributions have uncovered structure in this two-dimensional space, including correlations between component masses, mass ratio, and spins~\cite{2020ApJ...894..129S, 2021ApJ...922L...5C,2022PhRvD.105l3024F,2023ApJ...958...13A,2024A&A...692A..80P}, as well as evidence for multiple subpopulations~\cite{2024PhRvL.133e1401L,2023arXiv230401288G,2026ApJ...996...71H,2026PhRvL.137b1404P,2026PhRvL.137b1403B}.
Some of these features have been linked to distinct formation channels, such as stellar-collapse versus hierarchical-mergers~\cite{2017PhRvD..95l4046G,2021MNRAS.507.3362T,2024PhRvL.133e1401L}.

The joint distribution of mass and redshift also reveals how the population evolves over cosmic time, as several physical mechanisms predict a redshift-dependent mass distribution. 
The cosmic metallicity decreases toward earlier epochs, and metal-poor stars lose less mass to winds, leaving heavier remnants; hence, more massive BHs are expected to form at high redshift~\cite{2001A&A...369..574V, 2003ApJ...591..288H, 2010ApJ...714.1217B, 2014ARA&A..52..415M, 2018MNRAS.474.2959G, 2020MNRAS.498..495D, 2022ApJ...926...83T}. 
Moreover, different formation channels have distinct delay times, introducing an additional, complex redshift dependence~\cite{2022PhR...955....1M,2022MNRAS.515.5495M, 2022MNRAS.511.5797M}. 
For example, common-envelope evolution preferentially produces BHs below $\sim 30\,\Msun$ with short delay times, whereas the stable Roche-lobe overflow channel forms more massive systems with long delay times, making massive mergers relatively rare at high redshift~\cite{2022ApJ...931...17V}.  %
In dense star clusters, repeated mergers build heavy second-generation BHs that can fall inside the pair-instability mass gap~\cite{2016ApJ...831..187A, 2021NatAs...5..749G,2019MNRAS.486.5008A,2025PhRvL.134a1401A}. Because such mergers occur early in the host cluster's lifetime, the most massive products of this channel preferentially merge at high redshift~\cite{2024ApJ...967...62Y,2024A&A...688A.148T}.
These mechanisms thus predict distinct mass--redshift trends, so testing whether and how the mass distribution evolves with redshift offers a complementary probe of BBH formation channels.

Beyond probing formation channels, the mass--redshift dependence also carries implications for cosmological inference with GWs.
GW observations directly measure the luminosity distance to a source, while its redshift can be inferred statistically from the mass distribution, because the detector-frame masses are related to the source-frame masses by a redshift factor.
Leveraging this property, the ``spectral siren'' approach uses features of the intrinsic mass distribution, such as mass gaps and peaks, as reference scales that anchor the mass distribution in the source frame~\cite{2019ApJ...883L..42F,2021PhRvD.104f2009M,2022PhRvL.129f1102E}.
By tracking the detector-frame positions of these features as a function of luminosity distance, one can constrain cosmological parameters, most notably the Hubble constant $H_0$, without requiring electromagnetic counterparts or galaxy catalogs~\cite{2026arXiv260527227T}.
This method, however, relies on the assumption that the intrinsic mass features are stable. Any unmodeled redshift evolution of the mass distribution could bias the inferred cosmological parameters~\cite{2022MNRAS.515.5495M,2024PhRvD.109h3504P,2025ApJ...985..220T}. 
Thus, probing the mass--redshift dependence not only informs our understanding of BBH formation channels, but also helps quantify a potential systematic bias in cosmological parameters inferred from current GW spectral-siren measurements.

Many studies have investigated the redshift evolution of the BH mass distribution by reconstructing the conditional mass distribution as a function of redshift, using either parametric models with redshift-varying parameters or non-parametric approaches. 
For the smaller GWTC-2 sample, \citet{2021ApJ...912...98F} showed that the inferred evolution depends on the assumed mass distribution model. Modeling the primary-mass distribution as a power law with an abrupt high-mass cutoff, they found that the cutoff evolves with redshift, whereas a broken power-law model yielded no evidence for evolution with the same data.
For the larger GWTC-3 catalog, a parametric analysis found that the mass distribution could be fitted with a redshift-dependent peak, possibly associated with the pair-instability mass scale, although the Bayes factors remained inconclusive~\cite{2023MNRAS.523.4539K}. A fully non-parametric reconstruction of the same catalog reported evidence that the primary mass distribution evolves, with a subpopulation lighter than $\sim 20\,\Msun$ disappearing at $z > 0.4$~\cite{2024A&A...684A.204R}.
More recently, a non-parametric reconstruction of GWTC-3 and GWTC-4.0 found tentative evidence for a linear redshift evolution of the mass distribution above $\sim 50\,\Msun$~\cite{2025arXiv250925356A}.
On the other hand, several studies report no redshift evolution. \citet{2025A&A...698A..85L} found no evidence that either the $\sim 35\,\Msun$ feature or the power-law slope evolves below $z \sim 1$, and parametric models with a redshift- or time-dependent mass distribution likewise found no statistical support for evolution~\cite{2025PhRvD.111l3046G,2026arXiv260520112A}.
Non-parametric reconstructions at different redshifts also found no evidence for a change in the shape of the mass distribution~\cite{2023ApJ...957...37R,2025PhRvD.111f1305H,2026arXiv260911885I}.
Therefore, there is currently no consensus on whether such evolution exists.

An evolving mass distribution necessarily implies a statistical dependence between mass and redshift, and vice versa.
Testing for evolution in fact amounts to testing this dependence.
This calls for a tool that describes the dependence structure between two variables independently of their marginal distributions.
In this work, we employ the copula, a popular tool in modern statistics developed to describe complicated dependence structures between random variables, to probe the dependence between primary mass and redshift of BBH mergers in GWTC-5.0.
By Sklar's theorem, any joint distribution can be decomposed into its marginal distributions and a copula that fully encodes their dependence structure~\cite{Sklar1959,Nelsen2006}.
The mass--redshift dependence can thus be described entirely by the copula, while the marginal distributions are modeled separately.
Importantly, testing the dependence does not require assuming a parametric form for the conditional distribution of primary mass at a given redshift.
Moreover, different copula families encode different dependence structures, including asymmetric tail dependences.
Since different formation channels may produce mass--redshift dependences in different directions, this flexibility allows us to probe not only whether the mass distribution evolves, but also in which region of the mass--redshift plane the dependence is concentrated.
In GW astronomy, copulas have already been applied to BBH population analysis to measure the correlation between mass ratio and effective spin~\cite{2022MNRAS.517.3928A}, and were recently introduced to model the dependence structure between primary and secondary masses~\cite{2026arXiv260525980Q}.

\section{Copula}\label{sec:copula}


Consider two continuous random variables $x$ and $z$ with marginal cumulative distribution functions (CDFs) $F(x)$ and $W(z)$, respectively, and joint CDF $H(x,z)$.
Sklar's theorem guarantees the existence of a unique function $C$, called a `copula', defined on the unit square $[0,1]^2$ with uniform marginals, such that
\be
H(x,z|\theta_c) = C\left[F(x), W(z)| \theta_c\right],
\label{eq:jointcdf}
\ee
where $\theta_c$ is the parameter of the copula function $C$~\cite{Nelsen2006}.
Differentiating the joint CDF with respect to both variables yields the joint probability density function (PDF)
\be
h(x,z|\theta_c) = c\left[F(x), W(z)| \theta_c\right] f(x)\, w(z),
\label{eq:jointpdf}
\ee
where $f$ and $w$ are the marginal PDFs of $x$ and $z$, respectively, and
\be
c(u,v|\theta_c) \equiv \frac{\partial^2 C(u,v|\theta_c)}{\partial u\, \partial v}
\ee
is the density function of $C$, with $u\equiv F(x)$ and $v\equiv W(z)$ both uniformly distributed on $[0,1]$.
Equation~(\ref{eq:jointpdf}) shows that, once the marginal PDFs are specified, the joint PDF is fully determined by the copula density $c$. Different choices of the copula function encode different dependence structures between the two variables, independently of the marginal models.

Several families of copulas, each capable of describing different dependence structures, are commonly used in practice.
One class comprises copulas of elliptically contoured distributions, which are radially symmetric~\cite{Nelsen2006}. A representative member is the Gaussian copula, which corresponds to the dependence structure of a bivariate normal distribution. Its parameter $\theta_c\in(-1,1)$ is the correlation coefficient of that distribution.
Its CDF is
\be
C_{\rm G}(u,v|\theta_c) = \Phi_2\left[\Phi^{-1}(u), \Phi^{-1}(v)| \theta_c\right],
\ee
where $\Phi$ is the standard normal distribution CDF, $\Phi^{-1}$ its inverse, and $\Phi_2$ is the bivariate normal distribution CDF with unit variances and correlation $\theta_c$.
The corresponding density function is
\be
c_{\rm G}(u,v|\theta_c) = \frac{1}{\sqrt{1-\theta_c^2}}
\exp\!\left\{-\frac{\theta_c^2\left[\Phi^{-1}(u)\right]^2 + \theta_c^2\left[\Phi^{-1}(v)\right]^2 - 2\theta_c\,\Phi^{-1}(u)\,\Phi^{-1}(v)}{2\left(1-\theta_c^2\right)}\right\}.
\label{eq:cgauss}
\ee
Its parameter is related to Kendall's rank correlation coefficient by $\tau = (2/\pi)\arcsin\theta_c$, with $\theta_c=0$ corresponding to independence~\cite{Nelsen2006}.
Positive (negative) $\theta_c$ indicates a positive (negative) correlation between $x$ and $z$.


A second class is that of Archimedean copulas, constructed from a generator function $\phi$ of one variable via $C(u,v) = \phi^{-1}\!\left[\phi(u)+\phi(v)\right]$, where $\phi^{-1}$ denotes the inverse function of $\phi$. Different generators yield families with various dependence structures, many of which are asymmetric. Notable examples include the Clayton, Gumbel--Hougaard, and Frank copulas, which exhibit lower-tail dependence, upper-tail dependence, and radial symmetry, respectively~\cite{Nelsen2006}. Here, lower (upper) tail dependence means that the conditional probability that one variable takes an extremely small (large) value, given that the other does, remains nonzero in the limit~\cite{Nelsen2006}.

\begin{figure*}[t]
  \centering
  \includegraphics[width=0.99\columnwidth]{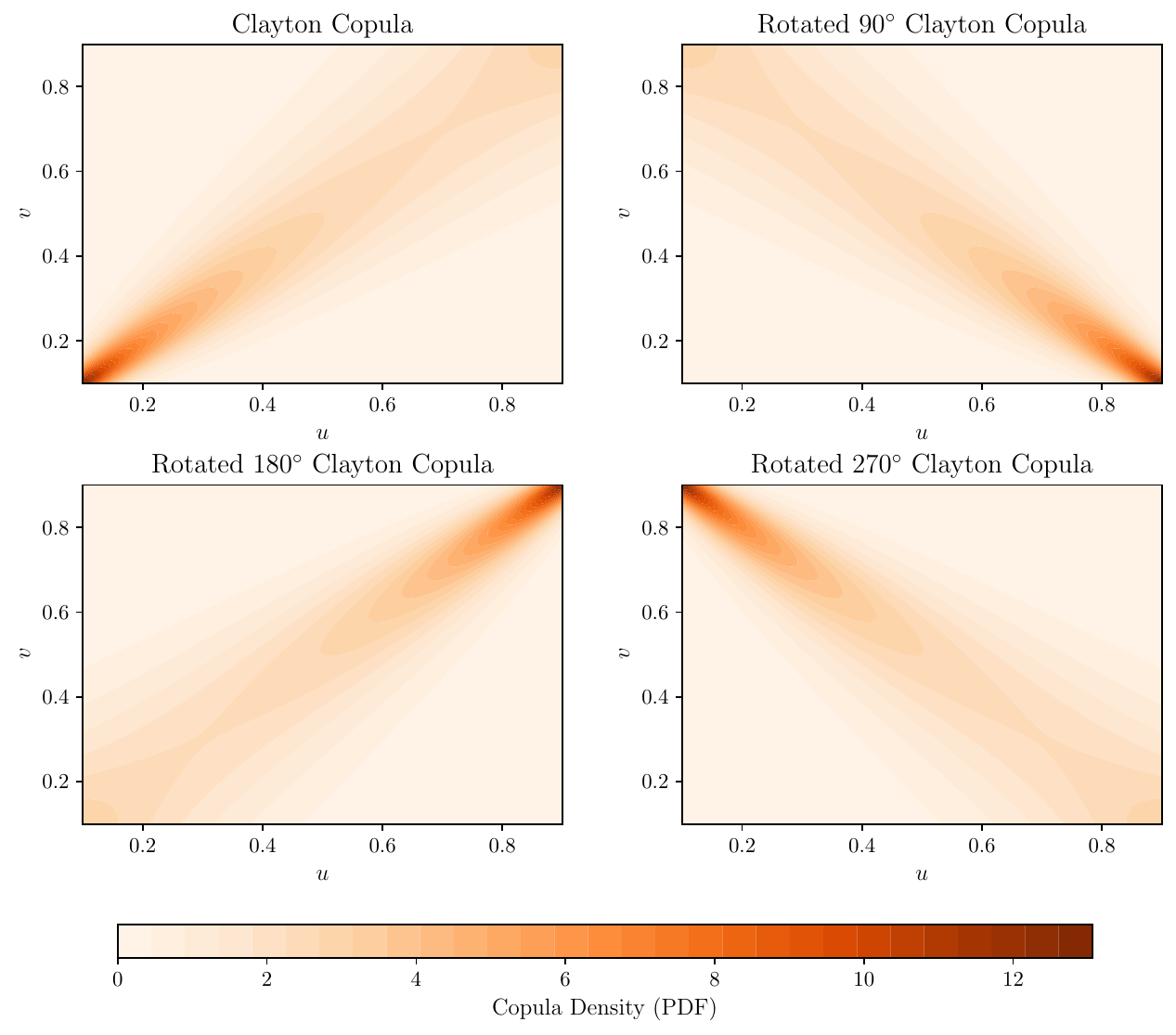}
  \caption{
	Densities of the Clayton copula and its three rotations with $\theta_c=5$.
    }
  \label{fig:clayton_rot}
\end{figure*}

The Clayton copula is obtained with the generator $\phi(t)=(t^{-\theta_c}-1)/\theta_c$, and its CDF is
\be
C_{\rm C}(u,v|\theta_c) = \left(u^{-\theta_c} + v^{-\theta_c} - 1\right)^{-1/\theta_c}, \qquad \theta_c>0,
\label{eq:claytoncdf}
\ee
with density
\be
c_{\rm C}(u,v|\theta_c) = (1+\theta_c)\,(uv)^{\theta_c}\left(u^{\theta_c}+v^{\theta_c}-(uv)^{\theta_c}\right)^{-2-1/\theta_c}.
\label{eq:cclayton}
\ee
The limit $\theta_c\to 0$ recovers the independence case, while $\theta_c\to\infty$ approaches perfect positive dependence; Kendall's tau is $\tau=\theta_c/(\theta_c+2)$~\cite{Nelsen2006}.
Unlike the Gaussian copula, which has no tail dependence in either tail, the Clayton copula has lower-tail dependence but no upper-tail dependence.
In the context of BBH population analysis, if $x$ denotes the primary mass and $z$ the redshift, this copula describes a mass--redshift dependence concentrated in the low-mass--low-redshift corner.
However, the form of the dependence between primary mass and redshift is not known a priori; it may be positive or negative, and may be concentrated in different corners in the mass--redshift plane.
Therefore, we also consider the three rotations of the Clayton copula in our analysis~\cite{Nelsen2006}.
The rotated densities are
\bea
c_{90}(u,v|\theta_c) &=& c_{\rm C}(1-u, v|\theta_c),\\
c_{180}(u,v|\theta_c) &=& c_{\rm C}(1-u, 1-v|\theta_c),\\
c_{270}(u,v|\theta_c) &=& c_{\rm C}(u, 1-v|\theta_c),
\eea
where the subscripts $90$, $180$, and $270$ denote the rotation angle in degrees.
The $180^\circ$ rotation shifts the tail dependence to the upper-right (high-mass--high-redshift) corner, concentrating dependence among massive, high-redshift binaries.
The $90^\circ$ and $270^\circ$ rotations instead describe negative dependence, concentrated among massive, low-redshift and light, high-redshift binaries, respectively.
The density values of these four models are illustrated in \Fig{fig:clayton_rot}.

In summary, our analysis considers five population models that differ only in the copula used to couple primary mass and the redshift. Hereafter, we refer to each model by its copula name. The Gaussian copula serves as a symmetric model without tail dependence; $\theta_c>0$ ($\theta_c<0$) indicates positive (negative) correlation between primary mass and redshift. The Clayton copula and its $90^\circ$, $180^\circ$, and $270^\circ$ rotations allow tail dependence to be concentrated in any of the four corners of the mass--redshift plane, \ie, low-mass--low-redshift, high-mass--low-redshift, high-mass--high-redshift, or low-mass--high-redshift. Each model is governed by a single dependence parameter $\theta_c$.
Additionally, we adopt as our baseline model the case of independence between primary mass and redshift, which corresponds to the Gaussian copula with $\theta_c=0$.

\section{Population inference}\label{sec:pop}

\subsection{The population model}\label{sec:popmodel}

We model the BBH population via the joint source-frame distribution of the primary mass $m_1$, the secondary mass $m_2$ (with $m_2\leq m_1$), and the redshift $z$.
Following \Eq{eq:jointpdf}, the differential event rate per unit detector-frame time $t_{\rm d}$, expressed in terms of the source-frame parameters $\vth_{\rm s}=(m_1, m_2, z)$, is written as
\be
\frac{\de N}{\de\vth_{\rm s} \de t_{\rm d}} \propto p(m_1, m_2, z|\bLa) = p(m_2|m_1, \bLa)\, p(m_1,z|\bLa),
\label{eq:jointpop}
\ee
where 
\be
p(m_1,z|\bLa)\equiv p(m_1|\bLa)\, p(z|\bLa)\, c\!\left[F_{m_1}(m_1), F_z(z)| \theta_c\right].
\ee
Here $\bLa$ denotes the set of hyperparameters, which includes both population and cosmological parameters; $F_{m_1}$ and $F_z$ are the CDFs of the primary mass and redshift, respectively; and $c$ is one of the copula densities described in \Sec{sec:copula}.
In this construction, the dependence between primary mass and redshift is introduced solely through the copula, while the marginal distributions are modeled independently using their commonly adopted forms. Therefore, no parametric assumption is required for the conditional distribution of the primary mass at a given redshift.
Note that the copula couples the redshift only to the primary mass, and the secondary mass becomes correlated with redshift indirectly through $m_1$.

For the primary mass distribution $p(m_1|\bLa)$, we adopt the \textsc{MultiPeak} model following \citet{2021ApJ...913L...7A}, which is a mixture of a power law and two Gaussian components:
\bea
p(m_1|\bLa) &\propto& S(m_1| \mmin, \delta_m)\Big[(1-\lambda_g)\, p_{\rm PL}(m_1| \mmin, \mmax, \alpha) \nn\\
&& + \lambda_g\lambda_{g,\rm low}\, G(m_1| \mu_{g,\rm low}, \sigma_{g,\rm low}, \mmin,\mmax) \nn\\
&& + \lambda_g(1-\lambda_{g,\rm low})\, G(m_1| \mu_{g,\rm high}, \sigma_{g,\rm high}, \mmin,\mmax)\Big],
\label{eq:pm1}
\eea
where $p_{\rm PL}(m|\mmin,\mmax,\alpha)\propto m^{-\alpha}$ is a power law with spectral index $\alpha$ between the minimum and maximum BH masses $\mmin$ and $\mmax$, and each $G(m|\mu_g,\sigma_g,\mmin,\mmax)$ is a Gaussian with mean $\mu_g$ and standard deviation $\sigma_g$, truncated and normalized on $[\mmin, \mmax]$.
The fraction $\lambda_g$ controls the total weight of the two peaks, and $\lambda_{g,\rm low}$ sets the relative weight of the low-mass peak.
The function
\be
S(m| \mmin, \delta_m) =
\begin{cases}
0, & m \leq \mmin,\\[4pt]
\left[1+\exp\!\left(\dfrac{\delta_m}{m-\mmin}+\dfrac{\delta_m}{m-\mmin-\delta_m}\right)\right]^{-1}, & \mmin < m < \mmin+\delta_m,\\[10pt]
1, & m \geq \mmin+\delta_m,
\end{cases}
\label{eq:smooth}
\ee
smooths the low-mass edge of the distribution over a mass scale $\delta_m$. 

For a fixed $m_1$, the secondary mass follows a smoothed truncated power law:
\be
p(m_2|m_1, \bLa) \propto S(m_2| \mmin, \delta_m)\, m_2^{\beta},
\qquad \mmin \leq m_2 \leq m_1,
\label{eq:pm2}
\ee
normalized to unity on the interval $[\mmin,\,m_1]$. The index $\beta$ governs the distribution of the secondary mass below $m_1$. 

For the redshift PDF $p(z|\bLa)$, we assume that the BBH merger rate density follows the cosmic star formation rate, parameterized with the Madau--Dickinson form~\cite{2014ARA&A..52..415M}
\be
R(z|\gamma, \kappa, z_p) \propto \left[1+(1+z_p)^{-(\gamma+\kappa)}\right]
\frac{(1+z)^{\gamma}}{1+\left(\dfrac{1+z}{1+z_p}\right)^{\gamma+\kappa}},
\label{eq:mdrate}
\ee
where $\gamma$ and $\kappa$ are the power-law slopes below and above the redshift turning point $z_p$, respectively. The normalization is chosen such that the shape factor equals unity at $z=0$.
The observed redshift distribution of mergers is then
\be
p(z\,|\,\bLa) \propto \frac{R(z|\bLa)}{1+z}\, \frac{\de V_c}{\de z}(z),
\label{eq:pz}
\ee
where the factor $1/(1+z)$ converts source-frame to detector-frame time and $\de V_c/\de z$ is the differential comoving volume.
We assume a flat $\Lambda$CDM cosmology, so $\de V_c/\de z = 4\pi c\, d_L^2(z)/[(1+z)^2 H(z)]$, with $c$ the speed of light, $d_L$ the luminosity distance, and $H(z)=H_0\sqrt{\Omega_{\rm m}(1+z)^3+1-\Omega_{\rm m}}$ the Hubble parameter, where $H_0$ is the Hubble constant and $\Omega_{\rm m}$ is the present-day matter density parameter.
We sample $H_0$ and fix $\Omega_{\rm m}$ to $0.3065$, which is the mean value inferred from Planck 2015~\cite{2016A&A...594A..13P}.

\subsection{Hierarchical likelihood}\label{sec:likelihood}

We infer the hyperparameters $\bLa$ with the hierarchical Bayesian framework for BBH population analyses~\cite{2019PASA...36...10T,2019MNRAS.486.1086M}, as implemented in the \textsc{icarogw} package~\cite{2023arXiv230517973M}.
Assuming that BBH mergers follow an inhomogeneous Poisson process, and adopting the scale-free prior on the expected number of detections (which discards information on the overall rate normalization), the hierarchical likelihood, evaluated via Monte Carlo integration using the posterior samples of the events, is
\be
\ln\mathcal{L}(\{\vd\}|\bLa) = \sum_{i=1}^{\Nobs}
\ln\!\left[\frac{1}{N_i^{s}}\sum_{j=1}^{N_i^{s}}
\frac{1}{\pi_{\rm PE}(\vth_{ij})}\,
\frac{\de N}{\de\vth_{\rm d}\de t_{\rm d}}(\vth_{ij}\,|\,\bLa)\right]
- \Nobs \ln \xi(\bLa),
\label{eq:likelihood}
\ee
where $\{\vd\}$ denotes the data of the detected events, $\Nobs$ is the number of detected events, $\vth_{ij}$ is the $j$-th of the $N_i^s$ posterior samples for the $i$-th event, and $\pi_{\rm PE}$ is the reference prior under which the parameter estimation samples were drawn.
Here $\vth_{\rm d}=(m_1^{\rm d}, m_2^{\rm d}, d_L)$ collects the detector-frame parameters. The differential event rate in the detector frame follows from \Eq{eq:jointpop} through the change of variables to detector-frame quantities:
\be
\frac{\de N}{\de \vth_{\rm d} \de t_{\rm d}}
= \frac{\de N}{\de\vth_{\rm s} \de t_{\rm d}} \frac{1}{|\det J_{\rm d\to s}|},
\label{eq:detrate}
\ee
with $m_{1,2}=m_{1,2}^{\rm d}/(1+z)$ and the Jacobian $|\det J_{\rm d\to s}|=(1+z)^2\, \de d_L/\de z$.

The detection fraction $\xi(\bLa)$, \ie, the fraction of events in the population expected to be detected, is estimated with a set of simulated signals (named injections) processed by the search pipelines~\cite{2023arXiv230517973M,2025PhRvD.112j2001E}:
\be
\xi(\bLa) = \frac{1}{N_{\rm gen}} \sum_{k=1}^{N_{\rm det}}
\frac{1}{p_{\rm draw}(\vth_k)}\,
\left.\frac{\de N}{\de\vth_{\rm d}\,\de t_{\rm d}}\right|_{k},
\label{eq:xi}
\ee
where $N_{\rm det}$ is the number of injections classified as detected, $N_{\rm gen}$ is the total number of injections generated, and $p_{\rm draw}$ is the distribution from which the injections were drawn.



We analyze the data with the five population models described in \Sec{sec:copula} plus the baseline model, defined as the Gaussian copula with $\theta_c=0$ held fixed.
The posterior distribution of the hyperparameters follows from the likelihood in \Eq{eq:likelihood} as
\be
p(\bLa\,|\,\{\vd\}) = \frac{\mathcal{L}(\{\vd\}|\bLa)\, \pi(\bLa)}{\mathcal{Z}},
\label{eq:posterior}
\ee
where $\pi(\bLa)$ is the prior on the hyperparameters. 
All hyperparameters are sampled with uniform priors, except $\Omega_{\rm m}$, which is fixed (see \Table{tab:priors}).
For the copula parameter we adopt $\theta_c\in\mathcal{U}(-0.9,0.9)$ for the Gaussian copula, and $\theta_c\in\mathcal{U}(0.01,10)$ for the Clayton copula and its rotations, spanning a range from nearly independent to strongly dependent distributions, with Kendall's $|\tau|\approx0.005$--$0.83$.
The normalization 
$
\mathcal{Z} \equiv \int \mathcal{L}(\{\vd\}|\bLa)\, \pi(\bLa)\, \de\bLa
$
in \Eq{eq:posterior} is the Bayesian evidence~\cite{2007MNRAS.378...72T,2008ConPh..49...71T}, which we use for subsequent model comparison. Parameter estimation is then performed by sampling the posterior distribution of \Eq{eq:posterior}.


\begin{table}[t]
\caption{\label{tab:priors}
Prior distributions of the hyperparameters. $\mathcal{U}(a,b)$ denotes a uniform distribution between $a$ and $b$.}
\begin{ruledtabular}
\begin{tabular}{ll}
Parameter & Prior \\
\hline
Hubble constant $H_0$ [$\kmsMpc$] & $\mathcal{U}(10, 200)$ \\
Matter density $\Omega_{\rm m}$ & $0.3065$ (fixed) \\
Primary mass power-law index $\alpha$ & $\mathcal{U}(1.5, 12)$ \\
Secondary mass power-law index $\beta$ & $\mathcal{U}(-4, 12)$ \\
Minimum BH mass $\mmin$ [$\Msun$] & $\mathcal{U}(2, 10)$ \\
Maximum BH mass $\mmax$ [$\Msun$] & $\mathcal{U}(50, 200)$ \\
Smoothing scale $\delta_m$ [$\Msun$] & $\mathcal{U}(0.001, 10)$ \\
Low-mass peak mean $\mu_{g,\rm low}$ [$\Msun$] & $\mathcal{U}(5, 100)$ \\
Low-mass peak standard deviation $\sigma_{g,\rm low}$ [$\Msun$] & $\mathcal{U}(0.4, 5)$ \\
High-mass peak mean $\mu_{g,\rm high}$ [$\Msun$] & $\mathcal{U}(5, 100)$ \\
High-mass peak standard deviation $\sigma_{g,\rm high}$ [$\Msun$] & $\mathcal{U}(0.4, 10)$ \\
Peak fraction $\lambda_g$ & $\mathcal{U}(0, 1)$ \\
Low-mass peak fraction $\lambda_{g,\rm low}$ & $\mathcal{U}(0, 1)$ \\
Low-redshift index $\gamma$ & $\mathcal{U}(0, 12)$ \\
High-redshift index $\kappa$ & $\mathcal{U}(0, 6)$ \\
Turning-point redshift $z_p$ & $\mathcal{U}(0, 4)$ \\
Copula parameter $\theta_c$ (Gaussian) & $\mathcal{U}(-0.9, 0.9)$ \\
Copula parameter $\theta_c$ (Clayton copula and its rotations) & $\mathcal{U}(0.01, 10)$ \\
\end{tabular}
\end{ruledtabular}
\end{table}

\section{Data}\label{sec:data}

We utilize the GWTC-5.0 catalog, which comprises 390 GW candidates detected up to the end of the O4b observing run~\cite{2026arXiv260527225T}.
We select events with a false alarm rate ${\rm FAR}\leq 0.25~{\rm yr}^{-1}$ and classify an event as a BBH when its median source-frame secondary mass satisfies $m_2>3\,\Msun$, thereby removing BNS and NS--BH systems.
We additionally exclude GW231123\_135430, whose inferred properties, such as binary masses and luminosity distance, exhibit a strong sensitivity to the waveform model employed in parameter estimation; the origin of this sensitivity remains unclear~\cite{2025ApJ...993L..25A,2026ApJ..1007L..17A,2026arXiv260527227T}.
The final sample consists of 235 BBH events.

For these selected events, we employ the public posterior samples released by the LVK collaboration~\cite{2026arXiv260527225T}. 
For events observed up to and including O3, we use the posterior samples generated with \textsc{IMRPhenomXPHM}~\cite{2021PhRvD.103j4056P}; for O4 events, we use those generated with \textsc{IMRPhenomXPHM-SpinTaylor}~\cite{2025PhRvD.111j4019C}.
From each event, we extract the posterior samples of the detector-frame binary masses $m_1^{\rm d}$, $m_2^{\rm d}$, and the luminosity distance $d_L$, which are used to evaluate the hierarchical likelihood in \Eq{eq:likelihood}.
The detection fraction $\xi$ in \Eq{eq:xi} is evaluated using the injection released by the LVK collaboration on Zenodo together with GWTC-5.0~\cite{2026arXiv260527226T,2025PhRvD.112j2001E}.
Consistent with the event selection criteria, we classify injections as detected if they satisfy ${\rm FAR}\leq 0.25~{\rm yr}^{-1}$ and have source-frame secondary mass $m_2>3\,\Msun$.

\section{Results}\label{sec:results}

\begin{figure}[tbp]
  \centering
  \includegraphics[width=0.495\columnwidth]{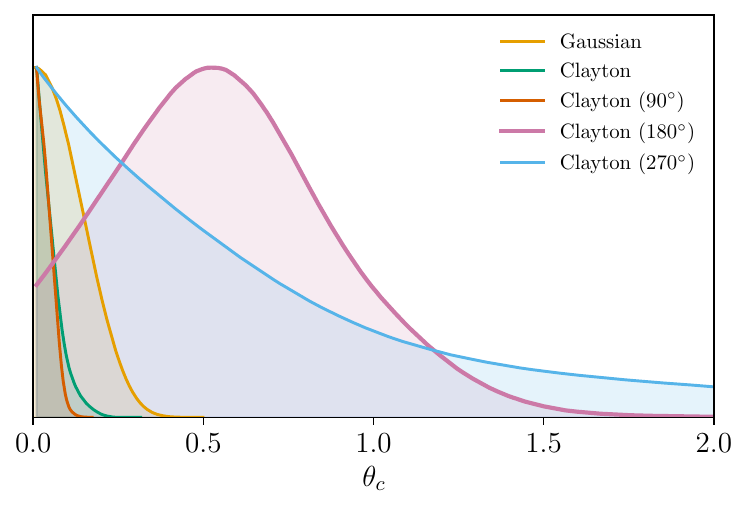}
  \includegraphics[width=0.495\columnwidth]{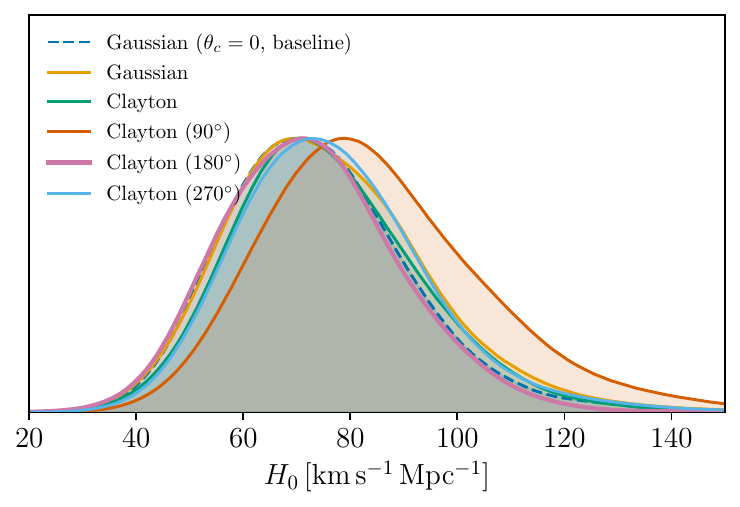}
  \caption{
	Marginalized one-dimensional posterior distributions of the copula parameter $\theta_c$ and of the Hubble constant $H_0$ for the six population models. 
    }
  \label{fig:constraints}
\end{figure}

The constraints on the copula parameter $\theta_c$ and on the Hubble constant $H_0$ for the six population models are listed in \Table{tab:results}, and the corresponding  marginalized one-dimensional posteriors are shown in \Fig{fig:constraints}. All quoted intervals correspond to the 68\% confidence level~(CL).

\begin{table*}[t]
\caption{\label{tab:results}
Constraints on the Hubble constant $H_0$ and the copula parameter $\theta_c$ for the six BBH population models, along with Kendall's $\tau$, the logarithmic Bayesian evidence, and the Bayes factor relative to the baseline model (defined in \Eq{eq:bayesfactor}). 
We show the posterior means and 68\% CL intervals. }
\small
\setlength{\tabcolsep}{3.5pt}
\begin{ruledtabular}
\begin{tabular}{lccccc}
Model & $H_0$ & $\theta_c$ & $\tau$ & $\ln\mathcal{Z}$ & $\ln\mathcal{B}$ \\
\hline
Gaussian ($\theta_c=0$, baseline) & $74.3_{-19.6}^{+13.7}$ & 0 (fixed) & 0 & $-638.93$ & 0 \\
Gaussian & $76.1_{-21.0}^{+14.8}$ & $0.01_{-0.12}^{+0.12}$ & $0.01_{-0.08}^{+0.08}$ & $-640.94$ & $-2.01$ \\
Clayton & $75.8_{-19.7}^{+14.3}$ & $<0.06$ & $<0.03$ & $-642.34$ & $-3.41$ \\
Clayton (rotated $90^\circ$) & $84.7_{-24.0}^{+15.9}$ & $<0.05$ & $<0.03$ & $-642.60$ & $-3.67$ \\
Clayton (rotated $180^\circ$) & $72.7_{-18.6}^{+14.3}$ & $0.58_{-0.38}^{+0.24}$ & $0.22_{-0.13}^{+0.07}$ & $-640.47$ & $-1.54$ \\
Clayton (rotated $270^\circ$) & $76.8_{-19.5}^{+14.3}$ & $<1.07$ & $<0.35$ & $-640.69$ & $-1.76$ \\
\end{tabular}
\end{ruledtabular}
\end{table*}

The Gaussian copula yields $\theta_c=0.01^{+0.12}_{-0.12}$, with Kendall's correlation coefficient $\tau=0.01_{-0.08}^{+0.08}$, fully consistent with a vanishing correlation between primary mass and redshift. For the Clayton copula and its $90^\circ$ and $270^\circ$ rotations, the posteriors of $\theta_c$ 
pile up at the lower boundary of the prior range, approaching the independence limit; we report the 68\% upper limits $\theta_c<0.06$, $\theta_c<0.05$, and $\theta_c<1.07$, respectively. In contrast, the $180^\circ$ rotated Clayton copula behaves differently: its $\theta_c$ posterior peaks at a nonzero value, $\theta_c=0.58^{+0.24}_{-0.38}$, corresponding to $\tau= 0.22_{-0.13}^{+0.07}$, which deviates from the independence limit $\theta_c \to 0$ by more than $1\sigma$. 
Relative to the baseline model, the five copula-based models shift the posterior mean of $H_0$ by varying amounts.
The largest shift occurs for the $90^\circ$ rotated Clayton copula, whose posterior mean exceeds the baseline value by $\sim 10~\kmsMpc$.
Notably, for the $180^\circ$ rotated Clayton model, the only case exhibiting a $>1\sigma$ preference for a nonzero $\theta_c$, the shift in the posterior mean of $H_0$ is only $\sim 1.6~\kmsMpc$, corresponding to approximately $2\%$ of the baseline value.
These shifts are small compared to the one-sided $68\%$ CL uncertainties of $13.7$--$24.0~\kmsMpc$ per model, so the $H_0$ constraints from the six models, whose posterior means range from $72.7$ to $84.7~\kmsMpc$, remain mutually consistent.
Therefore, with the current sample, the statistical uncertainty in $H_0$ still dominates over the systematic effect associated with the mass--redshift dependence within the models considered here.

\begin{figure*}[h]
  \centering
  \includegraphics[width=0.495\textwidth]{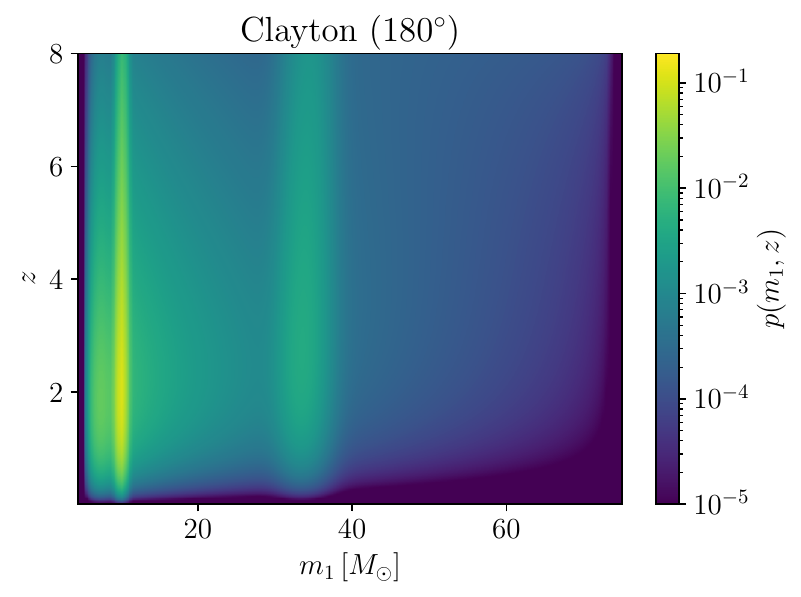}
  \includegraphics[width=0.495\textwidth]{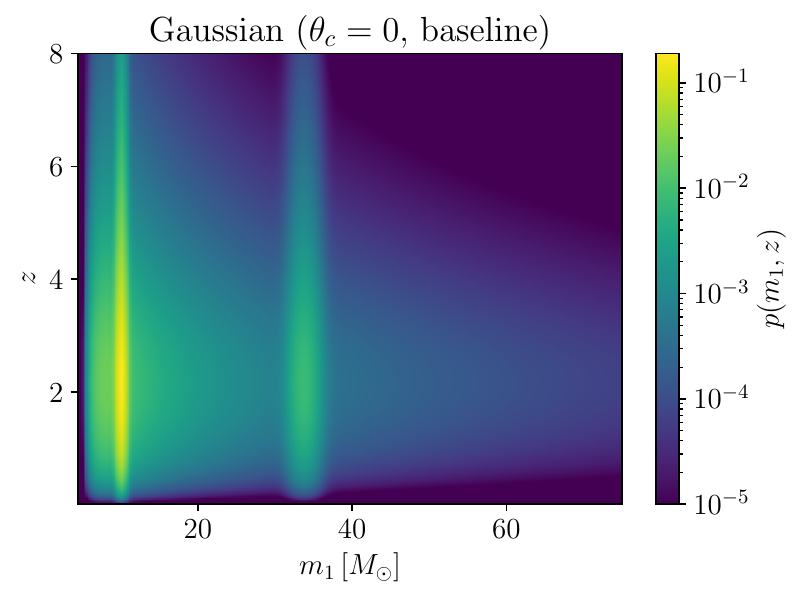}
  \caption{
	Joint source-frame distribution of primary mass and redshift, evaluated at the best-fit parameters of the $180^\circ$ rotated Clayton copula (left) and of the baseline model (right). 
    }
  \label{fig:pm1z}
\end{figure*}

\begin{figure}[h]
  \centering
  \includegraphics[width=0.495\columnwidth]{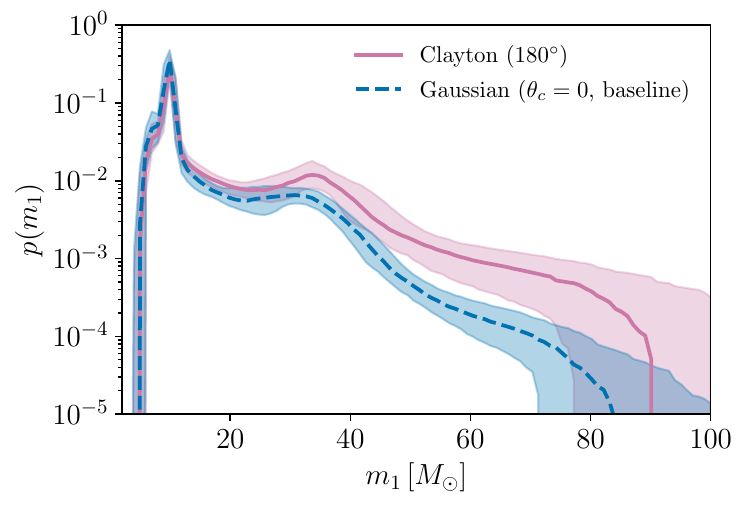}
  \includegraphics[width=0.495\columnwidth]{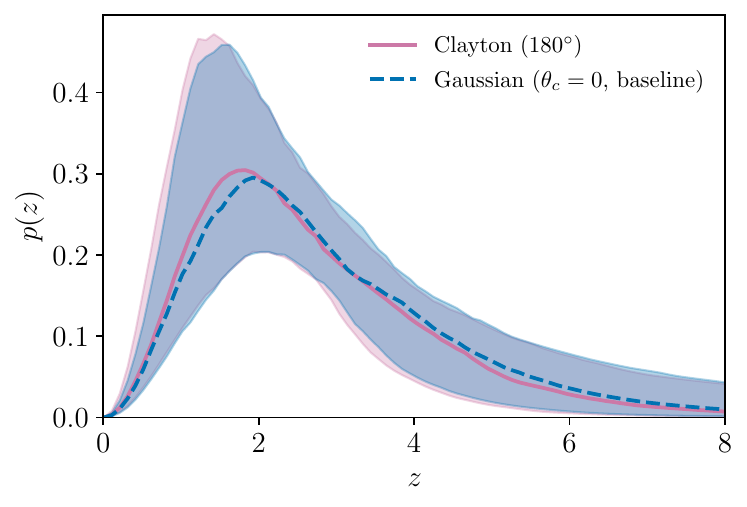}
  \caption{
	Marginalized primary mass distribution $p(m_1)$ (left panel) and redshift distribution $p(z)$ (right panel) for the $180^\circ$ rotated Clayton copula (solid) and the baseline model (dashed). The curves show the medians, and the shaded regions indicate the 68\% CL intervals.
    }
  \label{fig:marginals}
\end{figure}

The impact of the mass--redshift correlation on the population distribution is illustrated in \Fig{fig:pm1z}, which displays the joint distribution $p(m_1,z)$ evaluated at the best-fit parameters of the $180^\circ$ rotated Clayton copula and of the baseline model. Compared with the independent case, the $180^\circ$ rotated Clayton copula coupling visibly enhances the joint probability density of massive BHs at high redshift. However, the enhancement becomes less pronounced in the marginalized distributions displayed in \Fig{fig:marginals}. 
The redshift distribution $p(z)$ is nearly identical in the two models over the entire range shown. The primary mass distribution $p(m_1)$ also closely follows the baseline one at low mass, with only a slight separation above $m_1\sim 40\,\Msun$. 
This behaviour is expected, since the copula modifies only the dependence structure between the two variables, leaving the marginal distributions largely unchanged.

The mass--redshift dependence described by the $180^\circ$ rotated Clayton copula, in which massive primaries preferentially appear at high redshift, may be connected to several BBH formation scenarios. 
For example, more massive BHs may preferentially form from metal-poor progenitors~\cite{2010ApJ...714.1217B,2022ApJ...926...83T}, and the cosmic chemical enrichment makes such progenitors more abundant at early epochs.  %
Alternatively, massive BHs are assembled through repeated mergers in dense star clusters~\cite{2024ApJ...967...62Y,2024A&A...688A.148T}. Both naturally yield a higher prevalence of massive primaries at high redshift, precisely the region where the dependence is concentrated.

To assess whether the nonzero $\theta_c$ found for the $180^\circ$ rotated Clayton copula is statistically supported, we compare the six models through the Bayes factor, as reported in \Table{tab:results}. The Bayes factor of each copula-based population model relative to the baseline model is defined as
\be
\ln\mathcal{B}\equiv\ln\mathcal{Z}-\ln\mathcal{Z}_{\rm baseline},
\label{eq:bayesfactor}
\ee
and, according to the Jeffreys scale~\cite{Jeffreys1961,2008ConPh..49...71T}, $|\ln\mathcal{B}|<1$ is inconclusive, while $1<|\ln\mathcal{B}|<2.5$, $2.5<|\ln\mathcal{B}|<5$, and $|\ln\mathcal{B}|>5$ correspond to weak, moderate, and strong evidence against the model with the smaller $\ln\mathcal{Z}$, respectively. 
The baseline model attains the largest Bayesian evidence. All other copula models yield $\ln\mathcal{B}<0$, ranging from $\ln\mathcal{B}=-1.54$ for the $180^\circ$ rotated Clayton model to $\ln\mathcal{B}=-3.67$ for the $90^\circ$ rotated one. 
Thus, the evidence against the Gaussian copula and the $180^\circ$ and $270^\circ$ rotated Clayton models is weak, whereas that against the Clayton copula and its $90^\circ$ rotation is moderate.
Consequently, although the copula parameter of the $180^\circ$ rotated Clayton model deviates from zero by more than $1\sigma$, its statistical significance remains weak. The current BBH sample is therefore consistent with no redshift evolution of the primary mass distribution over the range covered by GWTC-5.0 ($z\lesssim1.2$), in agreement with the findings of Refs.~\cite{2023ApJ...957...37R,2025A&A...698A..85L,2025PhRvD.111l3046G,2025PhRvD.111f1305H,2026arXiv260520112A,2026arXiv260911885I}.

\section{Conclusion}\label{sec:5}

In this paper, we employ copula functions to investigate the dependence structure between the primary mass and the redshift of BBH mergers in the GWTC-5.0 catalog. A key advantage of this approach is that the copula fully describes the dependence between the two observables independently of their marginal distributions; consequently, it does not require assuming a parametric form for the conditional distribution of the primary mass at a given redshift. 
We construct the joint population distribution of primary mass, secondary mass, and redshift by coupling a \textsc{MultiPeak} primary-mass model and a Madau--Dickinson redshift distribution with either the Gaussian copula, the Clayton copula, or one of its $90^\circ$, $180^\circ$, and $270^\circ$ rotations. We then compare these five population models against a baseline model in which the primary mass and redshift are assumed independent.

Applying these models to 235 BBH events with the hierarchical Bayesian framework, we find that the Gaussian copula yields $\theta_c=0.01^{+0.12}_{-0.12}$ at the 68\% CL, fully consistent with a vanishing mass--redshift correlation. 
The posteriors of the Clayton copula and its $90^\circ$ and $270^\circ$ rotations pile up at the lower boundary of the prior range, giving the 68\% CL upper limits $\theta_c<0.06$, $0.05$, and $1.07$, respectively, also consistent with no mass--redshift dependence.
Notably, only the $180^\circ$ rotated Clayton copula exhibits a peak at a nonzero value of $\theta_c=0.58^{+0.24}_{-0.38}$, corresponding to Kendall's $\tau=0.22^{+0.07}_{-0.13}$ and exhibiting a $>1\sigma$ deviation from the independence limit ($\theta_c \to 0$). 
This dependence structure implies that massive primaries tend to appear preferentially at high redshift, a trend that may be connected to metal-poor progenitors, which are more abundant at early cosmic epochs, or to repeated mergers in dense star clusters. 

When comparing the models via Bayesian evidence, however, the baseline model attains the largest value. On the Jeffreys scale, the evidence against the Gaussian copula and the $180^\circ$ and $270^\circ$ rotated Clayton copulas is weak, whereas that against the Clayton and $90^\circ$ rotated Clayton models is moderate. 
Therefore, the current BBH sample remains consistent with no redshift evolution of the primary mass distribution over the range covered by GWTC-5.0 ($z\lesssim1.2$), in agreement with earlier studies~\cite{2023ApJ...957...37R,2025A&A...698A..85L,2025PhRvD.111l3046G,2025PhRvD.111f1305H,2026arXiv260520112A,2026arXiv260911885I}. 
Furthermore, the constraints on $H_0$ derived from the six models are mutually consistent within the $68\%$ CL. The baseline model gives $H_0=74.3^{+13.7}_{-19.6}~\kmsMpc$, indicating that the population models constructed here do not significantly affect the $H_0$ inference from the current dataset. The much larger BBH samples expected from next-generation GW detectors, such as the Einstein Telescope~\cite{2010CQGra..27s4002P} and Cosmic Explorer~\cite{2019BAAS...51g..35R}, will provide substantially more stringent tests of the mass--redshift dependence.

\begin{acknowledgments}
This work is supported by the Strategic Priority Research Program of the Chinese Academy of Sciences 
(grant No. XDB0550400),
the National Natural Science Foundation of China (grant Nos. 12603008, 12321003, 12422307, 12373053,
and 12275080), the China Manned Space Program (grant No. CMS-CSST-2025-A01), 
China Postdoctoral Science Foundation (grant No. 2025M783235),
and Postdoctoral Fellowship Program of CPSF (grant No. GZC20261721).
\end{acknowledgments}

\bibliography{ref_ADS}
\bibliographystyle{apsrev4-1}

\end{document}